# Directional fast neutron detection using a time projection chamber and plastic scintillation detectors

Yidong FU[1,2], Yang TIAN[1,2,*], Yulan LI[1,2], Jian YANG[1,2], Jin LI[1,2]

1. Dept. of Engineering Physics, Tsinghua University, Beijing 100084, China

2. Key Laboratory of Particle & Radiation Imaging (Tsinghua University), Ministry of Education, Beijing 100084, China

**Abstract** A new method for directional fast neutron detection is proposed based on a neutron time projection chamber (TPC) and position-sensitive plastic scintillation detectors. The detection system can efficiently locate the approximate location of a hot spot with 4π field-of-view using only the neutron TPC. Then, the system generates a high-resolution image of the hot spot using selected coincidence events in the TPC and the scintillation detectors. A prototype was built and tested using a $^{252}$Cf source. An efficiency of $7.1 \times 10^{-3}$ was achieved for fast searching. The angular resolution was 7.8 ° (full width at half maximum, FWHM) for high-resolution imaging using the simple back projection method.

## 1. Introduction

Fast neutron detection is an effective method for detecting special nuclear materials (SNMs). The direction of the SNM source can be further determined using fast neutron scatter imaging (Fig. 1). In the past decade, two types of detector systems have been designed for fast neutron scatter imaging and are namely the neutron scatter camera [1-2], and the neutron time projection chamber (TPC) [3-4].

* Corresponding author. E-mail address: yangt@mail.tsinghua.edu.cn (Yang TIAN).

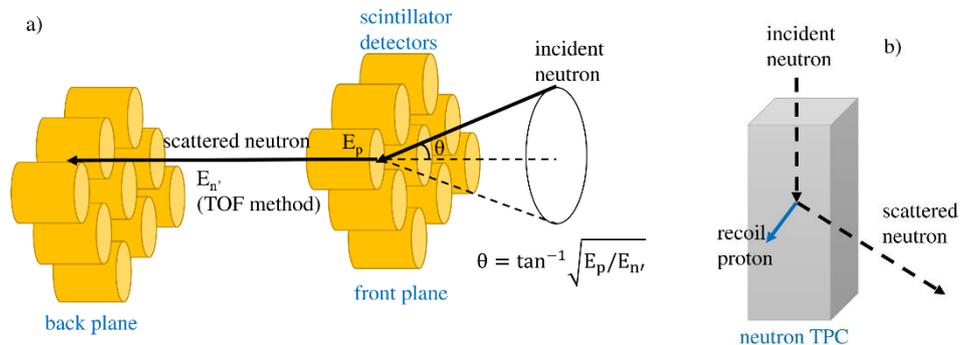

Fig. 1 Directional fast neutron detectors

(a) Neutron scatter camera; (b) Neutron TPC

The neutron scatter camera detects the direction of the fast neutrons based on a method similar to Compton imaging of γ-rays (Fig. 1 (a)) and has been utilized as a tool in the process of treaty verification and also in nuclear material monitoring [1]. A typical neutron scatter camera can achieve an angular resolution of 20° (full width at half maximum, FWHM) using the back projection method.

In a neutron TPC, the tracks of the recoil protons are reconstructed as shown in Fig. 1 (b) to determine the neutron source direction, by averaging the directions of a number of recoil protons [3]. The neutron TPC requires only a single scattering in the sensitive volume and is very efficient (from $10^{-3}$ to $10^{-2}$) with low gamma-ray background sensitivity.

This study incorporates position sensitive scintillation detectors to enhance the angular resolution of the neutron TPC [5]. As shown in Fig. 2, the effective event is second-order with the first scattering taking place in the neutron TPC which can record the recoil proton track, and with the scattered neutron then interacting with one of the plastic scintillation detectors. The scintillation detectors can provide the start time ($t_0$) for the TPC to measure the drift time and reconstruct the recoil proton track in the TPC. After the directions of the recoil proton and the scattered neutron are reconstructed, the direction of the incident neutron can be limited to one quadrant of a plane (the dashed area in Fig. 2). The effective event is selected based on the conclusion of the neutron-proton elastic scattering that the track of the scattered neutron is perpendicular to that of the recoil proton. The excellent track reconstruction ability of TPC results in a very high angular resolution using these coincidence events.

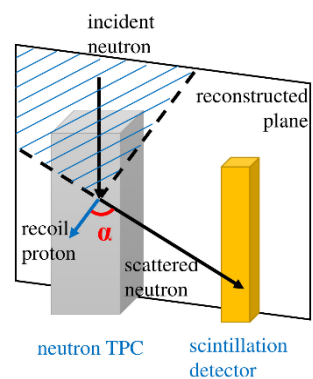

Fig. 2 Effective event and imaging principle.

A detector prototype was developed with a fast search mode to search for hot spots using only events in the neutron TPC. The fast search mode is efficient with $4\pi$ sensitivity. This prototype also has a high-resolution imaging mode which more accurately focuses on the hot spot using coincidence events in the neutron TPC and scintillation detectors. The detector design, simulation, and preliminary test results are presented herein.

## 2. System design

The prototype consisted of a neutron TPC and four position sensitive plastic scintillation detectors. The key parameters of the prototype are estimated by Monte Carlo simulation, and the results of the imaging system with more scintillation detectors are also included.

**2.1 Neutron TPC**

The neutron TPC shown in Fig. 3 was based on an existing nTPC [6] with the optimization performed on a readout pad design. It had a sensitive volume of 10×10×50 cm$^3$ with the signals recorded on a triple gaseous electron multiplier (GEM) detector with 4×4 mm$^2$ pads connected to 36 different 16-channel application specific integrated circuit (ASIC) boards [7]. Each channel was sampled by a 25 MHz flash analog to digital converter (FADC) [8]. Ar:C$_2$H$_6$ (50:50) was used as the working gas to achieve high efficiency.

The neutron TPC was tested using cosmic muons to calibrate the gain on each channel as shown in Fig. 4(a). A typical cosmic muon track is shown in Fig. 4(b).

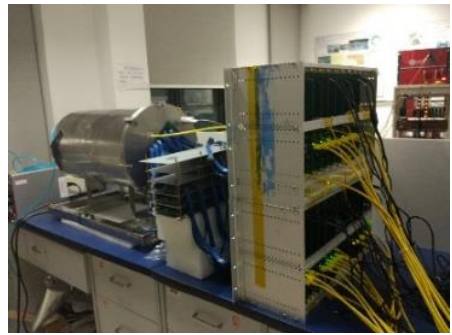

Fig. 3 Neutron TPC for the prototype system

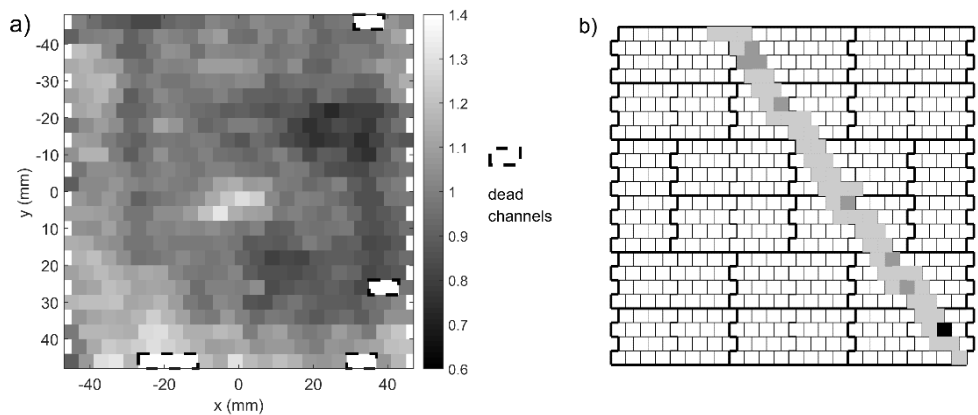

Fig. 4(a) Relative gains for all the channels. (b) Typical cosmic muon track is shown on the neutron TPC pad map

**2.2 Position sensitive plastic scintillation detectors**

The position sensitive plastic scintillation detectors each had a double-end photomultiplier tube

(PMT) readout. As shown in Fig. 5, each detector consisted of four SP101 plastic scintillator bars in a 2×2 array to achieve a better depth-of-interaction (DOI) resolution. This information was used to calculate the angle α in Fig. 2 to select the effective neutron events (α≈90°). Each scintillator bar was 1.6×1.6×40 cm$^3$. Tyvek paper was used as the diffuse reflector. The scintillation signal was outputted to two Hamamatsu CR105 PMTs and then sampled by a CAEN V1724 ADC. The DOI was calculated from the signal amplitudes of the two PMTs [5]. The detector was tested using a collimated $^{137}$Cs source which yielded a DOI resolution of 19 mm at 400 keV as shown in Fig. 6.

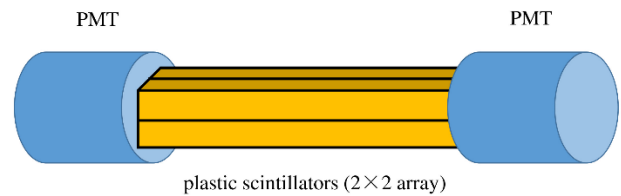

Fig. 5 Schematic diagram of a plastic scintillation detector

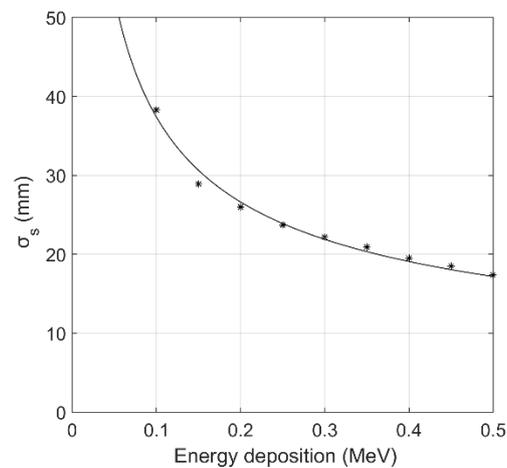

Fig. 6 DOI resolution of a plastic scintillation detector

## 2.3 Data acquisition system

The neutron TPC signal was used as the system trigger for the data acquisition system shown in Fig. 7, due to its good neutron/gamma discrimination ability. After one of the neutron TPC readout channels was triggered, the system then transmitted the scintillation detector signals that arrived during a time period equal to the width of the maximum drift time of the TPC to the computer. A LabView program was developed to control and record the data from both the neutron TPC and the scintillation detectors.

The coincidence neutron events for the high-resolution imaging mode were selected when the energy deposition of at least one scintillation detector exceeded a preset threshold during the recording time window. Then, the start time of the TPC drift time was determined by the scintillation detector signal, with the absolute positions in the axial direction of the triggered channels calculated from the drift time and the drift velocity of the gas. Finally, the directions of the recoil proton and the scattered neutron were calculated to determine if they were perpendicular to each other.

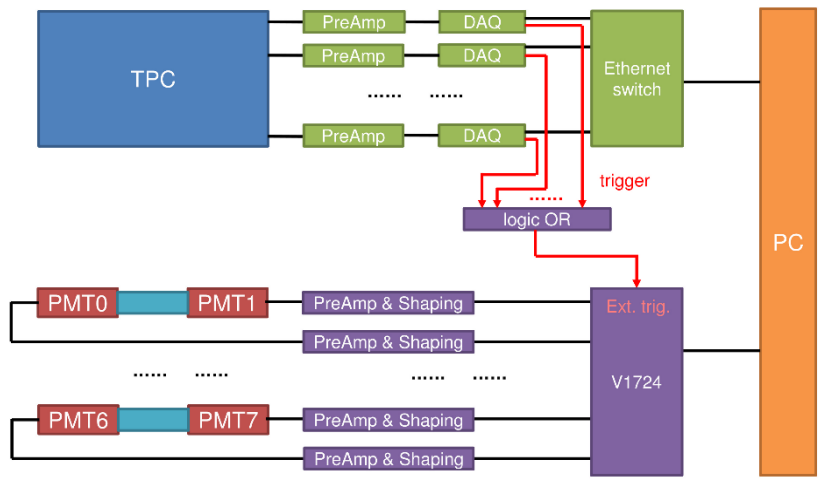

Fig. 7 Schematic diagram of the system data acquisition system

**2.4 Simulation results**

The key parameters of the high-resolution imaging mode were simulated for both the prototype system with 4 scintillation detectors and a system with additional scintillation detectors which covered the entire TPC chamber. The Monte Carlo simulation used Geant4, GARFIELD, and ROOT. The simulation setup is shown in Fig. 2. The simple back projection (SBP) algorithm was used for image reconstruction.

The simulated imaging result of the high-resolution mode is shown in Fig. 8. The simulation result shows that the addition of more scintillation detectors results in an increase of the detection efficiency (from $2.5\times10^{-5}$ to $2.7\times10^{-4}$) with no significant change in the angular resolution (7.0 ° for the prototype system and 7.1 ° for a system with 40 scintillation detectors, FWHM).

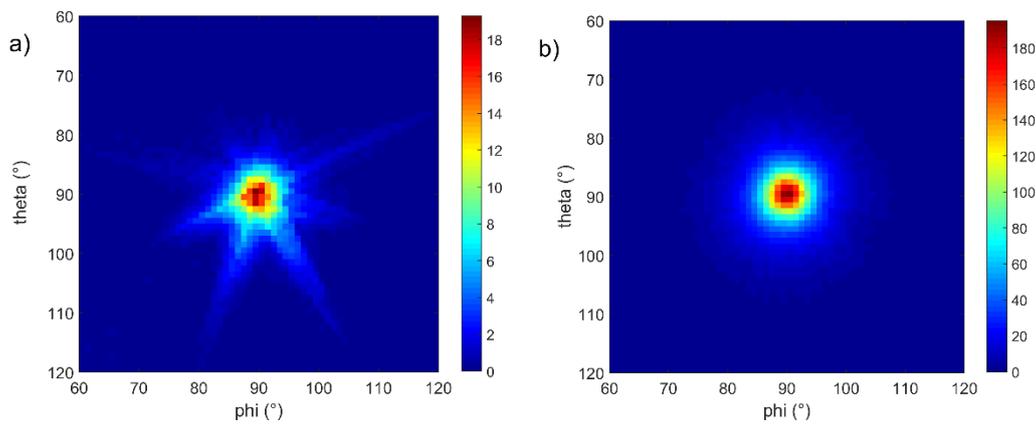

Fig. 8(a) Simulated image with 4 scintillation detectors;

(b) Simulated image with 40 scintillation detectors

## 3. Experimental results

The system was tested with a $^{252}$Cf neutron source in the China Institute of Atomic Energy. The four scintillation detectors were placed 21 cm from the center of the neutron TPC and were separated by 45 ° from each other (Fig. 9). The source was located 200 cm from the system.

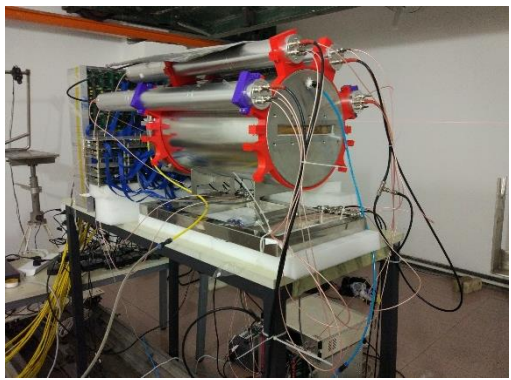

Fig. 9 The prototype system for the neutron test

**3.1 Fast searching mode**

Only the neutron TPC system was used for a fast search of the hot spots in a 4π field-of-view. In this mode, the direction of the incident fast neutron is determined by the reconstructed proton track as in a normal neutron TPC [3].

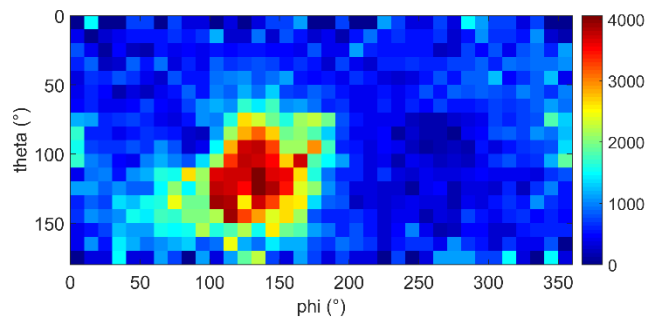

Fig. 10 Angular distribution of the recoil protons.

The result of the measurements is shown in Fig. 10. The angular resolution of the image is 91 ° (FWHM), which is a typical value of a neutron TPC. The efficiency is estimated to be $7.1 \times 10^{-3}$.

**3.2 High-resolution imaging mode**

After the directions of interest were determined using the fast search mode, high-resolution imaging was performed to more accurately locate the target using coincidence neutron events in the neutron TPC and the scintillation detectors. The system was rotated towards the hot spot to achieve better efficiency.

The coincidence neutron events were selected from the background by first using a time window equal to the maximum drift time in the neutron TPC. If there were both a proton track in the neutron TPC and an energy deposition event in a scintillation detector within the time window, then the angle α in Fig. 2 was calculated and compared with 90 °. The 90 ° criterion was used to efficiently distinguish the desired neutron events from background events.

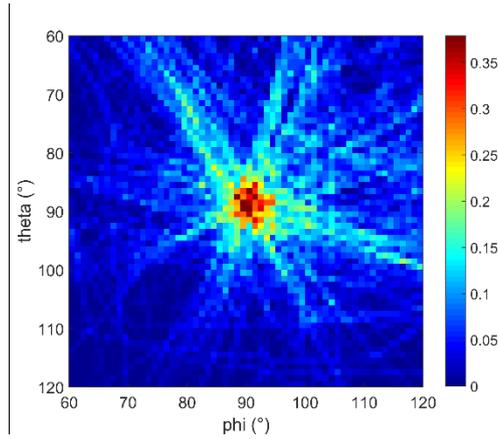

Fig. 11 SBP image of the high-resolution imaging mode.

Fig. 11 shows the neutron imaging result using the SBP method. The average angular resolution is 7.8 °(FWHM). The detection efficiency is estimated to be $2.2 \times 10^{-5}$.

## 4. Discussion

The neutron TPC has a very high detection efficiency, a large neutron/gamma discriminating ratio, and 4π field of view (FOV), which are very desirable for hotspot localization in neutron scatter imaging. With the aid of position-sensitive organic scintillation coincidence detectors, the angular resolution can be significantly enhanced (from 91 ° to 7.8 ° FWHM). The feasibility was testified using the prototype, which had an angular resolution which was even better than that of a typical neutron scatter camera. To the best of our knowledge, this is the highest reported resolution among the existing devices based on the SBP method. A high angular resolution is very important for the detailed measurement of the distribution within a hot spot, which can be further enhanced by using improved reconstruction methods such as filtered back projection (FBP) or the maximum likelihood estimation method (MLEM).

There is no need for energy measurement when imaging with the TPC and scintillator coincidence events. As a result, the relatively large uncertainty in the angular measurement introduced by the organic scintillator due to a poor energy resolution can be avoided. The limited efficiency can be improved by using more organic scintillation detector modules. The efficiency is also limited by the DOI resolution of the scintillation detectors. Due to the excellent angular resolution of the TPC [6], the resolution of angle α is mainly determined by the DOI resolution. When angle α has a better resolution, more neutron events can be distinguished from the accidental coincident background to improve the efficiency within a given window at 90°(85°– 95°in the test). By increasing the number and improving the DOI resolution of the scintillation detectors, a similar efficiency to a typical neutron scatter camera can be obtained.

One disadvantage of the prototype is the relatively complex setup of the neutron TPC (e.g., gas flow, instability of the amplifying stage, complex data acquisition system), which currently renders it unsuitable for field use. The key component of the TPC is a micro pattern gaseous detector (MPGD) with 2-D pad readout. Some novel MPGD components have been optimized for industrial production and field application such as the PTFE THGEM [9] and the glass GEM [10], whose stability, robustness, and compatibility for vacuum and sealing operation have been improved. The cross strip 2-D readout has already been used in a TPC [3], which can simplify the design of the DAQ. The aforementioned improvements can assist in optimizing the future design of the neutron TPC for field use.

## 5. Conclusion

Directional fast neutron detection is effective in SNM detection. High angular resolution, reasonable efficiency, and 4π FOV are important and desirable in practical applications. The achievement of these characteristics continues to be a challenge for existing devices. We propose a novel method based on a time projection chamber and plastic scintillation detectors to address these limitations.

The system initially efficiently searches for a hotspot in a 4π field-of-view, then produces a high-resolution image of the hot spot. The feasibility of imaging a fission neutron source was experimentally demonstrated. The efficiency and angular resolution of the prototype can be further improved by using more scintillator modules and by exploiting improved reconstruction methods. We suggest that the combined TPC and scintillator system is suitable for SNM detection.

**Acknowledgments**

We sincerely thank Yina Liu and Hailong Luo of the China Institute of Atomic Energy for useful discussions about the neutron test.